\documentstyle[twocolumn,prl,aps,floats,epsf]{revtex}

\begin{document}
\draft

\title{Meron Pairs and Fermion Zero Modes}

\author{James V.\ Steele and John W.\ Negele}
\address{Center for Theoretical Physics, Laboratory for Nuclear
Science, and Department of Physics\\Massachusetts Institute of
	Technology, Cambridge, MA\ \ 02139, USA}

\date{\today}

\twocolumn[\hsize\textwidth\columnwidth\hsize\csname@twocolumnfalse\endcsname

\maketitle

\begin{abstract}
Merons, conjectured as a semiclassical mechanism for color
confinement in QCD, have been described analytically by 
either infinite action configurations or an {\it Ansatz} with
discontinuous action.
We construct a smooth, finite action, stationary lattice solution
corresponding to a meron pair.
We also derive an analytical solution for the zero mode of the 
meron pair {\it Ansatz}, show that it has the
qualitative behavior of the exact zero mode of the lattice solution,
and propose the use of zero modes to identify meron gauge field
configurations in stochastic evaluations of the lattice QCD path
integral. 
\end{abstract}

\pacs{PACS number(s): 12.38.Aw, 11.15.Ha, 11.15.Kc, 12.38.Gc}

]



Chiral symmetry breaking and confinement of color charge,
the two essential features of low-energy QCD, cannot be understood
perturbatively in the coupling constant.
Two complementary nonperturbative approaches are 
analytical approximation using
semiclassical analysis of the QCD path integral
(analogous to the WKB approximation in quantum mechanics) 
and numerical solution using lattice gauge theory.
Combining physical understanding from 
the semiclassical approximation
with systematic rigor of the lattice 
has provided valuable insight into hadron structure~\cite{Chu:1994vi}
and our goal is to use this combined approach to study confinement.

Most semiclassical treatments of QCD have focused on instantons,
leading to a qualitative, and in some cases even quantitative,
understanding of chiral symmetry breaking and the structure of the QCD
vacuum~\cite{Schafer:1998wv}.
However, simple configurations of instantons, such as a dilute
gas~\cite{Callan:1977qs} or a random superposition~\cite{Chen:1999ct}, 
do not confine color charges.
Of alternative semiclassical mechanisms
proposed for 
confinement, including merons~\cite{Callan:1977qs},
monopoles~\cite{Smit:1991vg}, and center 
vortices~\cite{Gonzalez-Arroyo:1996zy},
we focus on merons,
which are solutions to the classical Yang-Mills field equations
with one-half unit of localized topological charge.
They are the (3+1)-dimensional nonabelian 
extension of the 't~Hooft--Polyakov monopole
configuration, which has been shown to lead to confinement in 2+1
dimensions~\cite{Polyakov:1977fu}. 
Ultimately, we seek to understand if merons generate area and
perimeter laws characterizing the behavior of confined and screened
color sources.

A purely analytic study of merons has proven intractable
for several reasons.
Unlike instantons, no exact solution exists for more than two
merons~\cite{Actor:1979in}, 
gauge fields for isolated merons fall off too slowly ($A\sim1/x$)
to superpose them, and the field strength is singular.
A patched {\it Ansatz} configuration that removes the
singularities~\cite{Callan:1977qs} does not satisfy the classical
Yang-Mills equations, preventing even
calculation of 
gaussian fluctuations around 
a meron pair~\cite{Laughton:1980vv}.

This letter lays the groundwork for using lattice techniques to
investigate the presence and role of merons in QCD.
We construct a smooth, finite action configuration
corresponding to a meron pair, which with its associated
quantum fluctuations, should be present in 
Monte Carlo lattice QCD calculations.
Similar work with instantons has shown fermionic zero
modes are a powerful tool for identifying topological objects on the
lattice.
Therefore, we also analytically derive the zero mode of the continuum
meron pair {\it Ansatz} and compare with exact numerical results.
For simplicity, we restrict our discussion to SU(2) color without loss
of generality.


Two known solutions to the classical Yang-Mills
equations in four Euclidean dimensions
are instantons and merons.
Both have topological charge, can be interpreted as tunneling
solutions, and can be written in the general form 
(for the covariant derivative $D_\mu = \partial_\mu - i A_\mu^a
\sigma^a/2$  and 't~Hooft symbol $\eta_{a\mu\nu}$)
\begin{equation}
A_\mu^a(x) =  -\eta_{a\mu\nu} \partial_\nu \ln \Pi(x^2)
\ ,
\label{general}
\end{equation}
with $\Pi(x^2) = 1/(x^2+\rho^2)$ for an instanton
and $\Pi(x^2)=1/\sqrt{x^2}$ for a meron.

Conformal symmetry of the classical Yang-Mills action, in
particular under inversion $x_\mu\to x_\mu/x^2$, shows that
in addition to a meron at the origin, there is 
a second meron at infinity. 
These two merons can be mapped to arbitrary positions, which we define
to be the origin and $d_\mu$.
After a gauge transformation, the gauge field for the two merons
takes the simple form~\cite{deAlfaro:1976qz}
\begin{equation}
A_\mu^a(x) =  \eta_{a\mu\nu} 
\left[ \frac{x^\nu}{x^2}+ \frac{(x-d)^\nu}{(x-d)^2} \right] \ .
\label{conform}
\end{equation}
Similar to instantons~\cite{Schafer:1998wv}, a meron pair can be
expressed in 
singular gauge by performing a large gauge transformation
about the mid-point of the pair, resulting in a gauge field
that falls off faster at large distances ($A\sim x^{-3}$).
A careful treatment of the singularities shows that the
topological charge density is~\cite{deAlfaro:1976qz}
\begin{equation}
Q(x) \equiv 
\frac{\mbox{Tr}}{16\pi^2} 
 \left( F_{\mu\nu} \widetilde F^{\mu\nu} \right)
 = \frac12\, \delta^4(x) + \frac12\, \delta^4(x-d) \ ,
\label{topch}
\end{equation}
yielding total 
topological charge $Q=1$, just like the instanton.

The gauge field Eq.~(\ref{conform}) has infinite action density 
at the singularities $x_\mu=\{0,\, d_\mu\}$.
Hence, a finite action {\it Ansatz} has been suggested~\cite{Callan:1977qs}
\begin{equation}
A_\mu^a(x) =  \eta_{a\mu\nu} x^\nu \left\{ 
\begin{array}{cll}
\displaystyle
\frac{2}{x^2+r^2} \ ,\qquad & \sqrt{x^2} < r \ ,
\\[3ex]
\displaystyle
\frac{1}{x^2} \ ,\qquad & r < \sqrt{x^2} < R \ ,
\\[3ex]
\displaystyle
\frac{2}{x^2+R^2} \ ,\qquad & R < \sqrt{x^2} \ .
\end{array}
\right.
\label{caps}
\end{equation} 
Here, the singular meron fields for $\sqrt{x^2}<r$ and $\sqrt{x^2}>R$
are replaced by instanton caps, each containing topological charge 
$\frac12$ to agree with Eq.~(\ref{topch}).
The radii $r$ and $R$ are arbitrary and we
will refer to the three separate regions defined in
Eq.~(\ref{caps}) as regions I, II, and III, respectively.
The action for this configuration is (with $S_0=8\pi^2/g^2$)
\begin{equation}
S= S_0 \left( 1 + \frac38 \ln \frac{R}{r} \right) \ ,
\label{action}
\end{equation}
which shows the divergence in the $r\to0$ or $R\to\infty$ limit.
Although this patching of instanton caps is continuous,
the derivatives are not, and so the equations of
motion are violated at the
boundaries of the regions in Eq.~(\ref{caps}).

\begin{figure}[t]
\vspace*{-1cm}
\begin{center}
\leavevmode
\epsfxsize=3.6in
\epsffile[170 608 460 720]{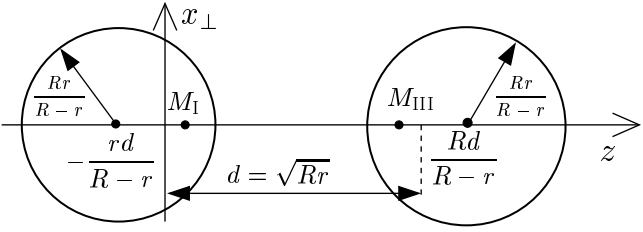}
\end{center}
\caption{\label{fig1} Meron pair separated by $d=\sqrt{R r}$ regulated
with instanton caps, each containing $\frac12$ topological charge.
} 
\vspace{-.4cm}
\end{figure}

Applying the same transformations used to attain Eq.~(\ref{conform})
to the instanton cap solution, regions I and III become
four-dimensional spheres 
each containing half an instanton~\cite{Callan:1977qs}.
The geometry of the instanton caps
is shown in Fig.~\ref{fig1} for the symmetric choice
of displacement $d=\sqrt{R r}$ along the $z$-direction.
Note that the action for the complicated geometry of
Fig.~\ref{fig1} is still given by Eq.~(\ref{action}). 
We will concentrate on this case below, and generalization to a different
$d$ is straightforward.

The original positions of the two merons $x_\mu=\{0,\,d_\mu\}$
are not the centers of the spheres, nor are they the positions of
maximum action density, which occurs within the spheres  
with $S_{\rm max} = (48/g^2) (R+r)^4/d^8$ at
\begin{equation}
\left(M_{\rm I}\right)_\mu = \frac{r^2}{r^2+d^2} \; d_\mu \ ,
\qquad
\left(M_{\rm III}\right)_\mu = \frac{R^2}{R^2+d^2} \; d_\mu \ . 
\label{maxima}
\end{equation}
However, in the limit $r\to0$ and
$R\to\infty$ holding $d$ fixed, the spheres will shrink around the
original points reducing to the bare meron pair in Eq.~(\ref{conform}). 
In the opposite limit $R\to r$, the radii of the spheres
increase to infinity, leaving an instanton of size $\rho = d$.

An instanton can therefore be interpreted as consisting of a meron
pair. 
This complements the fact that instantons and antiinstantons have
dipole interactions between each other.
If there exists a regime in QCD where meron entropy contributes more
to the free energy than the logarithmic potential between pairs, like
the Kosterlitz--Thouless phase transition, 
instantons could break apart into meron
pairs~\cite{Callan:1977qs}. 
The intrinsic size of the instanton caps would then be determined by the
instanton scale $\rho$.  


The Atiyah-Singer index theorem states that 
a fermion in the presence of 
a gauge field with topological charge $Q$, described by the equation
\begin{equation}
\raise.15ex\hbox{$/$}\kern-.65em\hbox{$D$}
\, \psi = \lambda \psi \ , \quad \mbox{with} \quad
\psi = \left( {\psi_R \atop \psi_L} \right) \ ,
\label{dirac}
\end{equation}
has $n_R$ right-handed and $n_L$ left-handed
zero modes (defined by $\lambda=0$) such that $Q=n_L-n_R$.
This can sometimes be
strengthened by a vanishing theorem~\cite{Jackiw:1977pu}.
Applying $\raise.15ex\hbox{$/$}\kern-.65em\hbox{$D$}$
twice to $\psi$ decouples the right- and left-hand
components, and focusing on the equation for $\psi_R$, gives
\begin{equation}
\left( D^2 + \frac12 \bar\eta_{a\mu\nu} \sigma^a F^{\mu\nu} \right) \psi_R 
=0\ ,
\label{van}
\end{equation}
where $\sigma^a$
acts on the spin indices of $\psi_R$ and $F^{\mu\nu}$ 
acts on the color indices.
For a self-dual gauge field (like an instanton), the second term in
Eq.~(\ref{van}) is zero; and since $D^2$ is a negative definite
operator, there are no 
normalizable right-handed zero modes, implying $Q=n_L$.
Although the meron pair is not self-dual, the second term can be shown
to be negative definite as well, leading to the same conclusion.

The finite action $Q=1$ gauge field Eq.~(\ref{general})
has a fermion zero mode given by
\begin{equation}
\psi = \left( {0 \atop \phi} \right) \ , \qquad
\phi_\alpha^a = {\cal N} \;  \Pi^{3/2} \;
\varepsilon_\alpha^a \ ,
\label{zmode}
\end{equation}
with normalization ${\cal N}$ and
$\varepsilon = i \sigma_2$ coupling the spin index $\alpha$ to the
color index $a$ (both of which can be either 1 or 2) in a singlet
configuration.
The gauge field for the meron pair with instanton caps corresponds to
\begin{equation}
\!\!\!\Pi(x) = \left\{
\begin{array}{ll}
\displaystyle
\frac{2 \xi_i d^2}{x^2 + \xi_i^2 (x-d)^2}\ , 
& \mbox{for regions $i=$ I, III,}
\\[3ex]
\displaystyle
\frac{d^2}{\sqrt{x^2 (x-d)^2}} \ , 
& \mbox{region II,} 
\end{array}
\right.
\end{equation}
with $\xi_{\rm I}=r/d$ and $\xi_{\rm III}=R/d$.
The normalized solution to the zero mode is then Eq.~(\ref{zmode})
with 
$
{\cal N}^{-1} = 2\pi d^2 \left[2-\sqrt{r/R} \right]^{1/2} .
$
Note that the unregulated meron pair ($r\to0$, $R\to\infty$)
has a normalizable zero mode itself~\cite{Konishi:1999re} 
\begin{equation}
\phi_\alpha^a(x) = \frac{d\; \varepsilon_\alpha^a}
{2\sqrt{2} \pi \left( x^2 (x-d)^2 \right)^{3/4}} \ .
\end{equation}
The gauge-invariant zero mode density $\psi^\dagger \psi(x)$ 
has a bridge between two merons that falls off like $x^{-3}$
in contrast to the $x^{-6}$ fall-off in all other directions.
This behavior can be used to identify merons when analyzing their zero
modes on the lattice, similar to what was done for instantons in
Ref.~\cite{Ivanenko:1998nb}. 


As mentioned above, the patching of instanton caps to obtain an
explicit analytic solution has unavoidable and unphysical
discontinuities in the action density.
Therefore, 
in order to study this solution further,
we put the gauge field on a space-time lattice of spacing $a$ in a
box of size $L_0\times L_1\times L_2\times L_3$.
The gauge-field degrees of freedom
are replaced by the usual parallel transport link variable,
$U_\mu(x) = {\rm P} \exp \left[-i \int_x^{x+a\, \hat e_\mu} 
A_\nu(z) dz^\nu\right]$.
The exponentiated integral can be performed analytically
within the instanton caps, producing arctangents.
Outside of the caps, 
the integral is evaluated numerically by 
dividing each link
into as many sub-links as necessary to reduce the 
${\cal O}(a^3)$ path-ordering errors below machine precision.

\begin{figure}
\vspace*{-.5cm}
\begin{center}
\leavevmode
\epsfxsize=3.2in
\epsffile{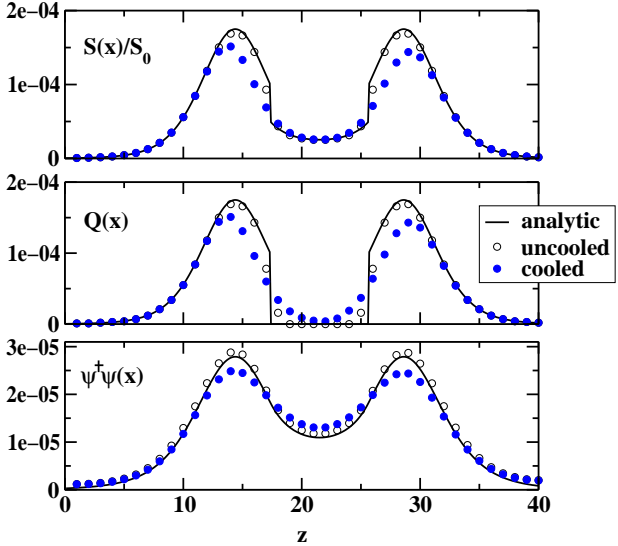}
\end{center}
\caption{\label{fig2} The action density $S(x)$ for the
regulated meron pair,
normalized by the instanton action $S_0=8\pi^2/g^2$, 
sliced through the center of the configuration in the direction of
separation. 
Shown are the analytic (solid), initial lattice (open circles)
and cooled lattice (filled circles) configurations.  The same is shown
for the topological charge density $Q(x)$ and fermion zero mode
density $\psi^\dagger\psi(x)$.}
\vspace{-.35cm}
\end{figure}

\begin{figure}[t]
\vspace*{-.8cm}
\begin{center}
\leavevmode
\epsfxsize=3.6in
\hbox{\hspace*{-.1in}\epsffile{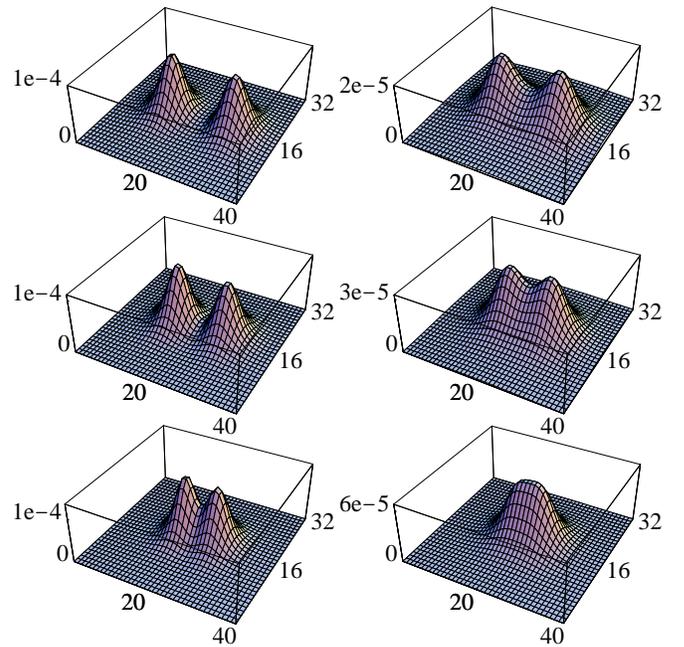}}
\end{center}
\vspace{-.35cm}
\caption{\label{fig3} The topological charge density $Q(x)$ on the
left and
fermion zero mode density $\psi^\dagger\psi(x)$ on the right
for a cooled meron pair
with $r=9$, separated by distance $d=20.8$, $17.7$, and $14.1$ 
respectively,
in the $(z,t)$-plane. }
\vspace{-.35cm}
\end{figure}

We then calculate the action density $S(x)$ for a meron pair with
instanton caps using the improved action of
Ref.~\cite{GarciaPerez:1994ki} and 
the topological charge density $Q(x)$ using 
products of clovers.
These (open circles) are compared with the patched {\it Ansatz}
results (solid line)
in Fig.~\ref{fig2} for $r=9$ and $d=20.8$ 
(in units of the lattice spacing) on a $32^3\times40$ lattice.
We use the Arnoldi method to solve for the zero mode of this
gauge configuration and hence the density $\psi^\dagger \psi(x)$,
which is also compared with the analytic result in the same figure,
showing excellent agreement.

Two important features of the patched {\it Ansatz} also evident with the
lattice representation are the discontinuities in the action density 
at the boundary of the instanton caps and
vanishing of the topological charge density outside of the caps as given
by Eq.~(\ref{topch}).
The discontinuity in the action density is unphysical and can be 
smoothed out by using a
relaxation algorithm to iteratively minimize the lattice
action~\cite{Berg:1981nw}.
On a sweep through the lattice, referred to as a cooling step,
each link is chosen to locally minimize the action density.
Since a single instanton is already a minimum of the action, 
this algorithm would leave the instanton unchanged (for a suitably
improved lattice action).
The result for the regulated meron pair
after ten cooling steps is represented in Fig.~\ref{fig2} by the
filled circles, showing the discontinuities in the action density
have already been smoothed out.
In Fig.~\ref{fig3}, we 
also plot side-by-side the cooled $Q(x)$ and $\psi^\dagger \psi(x)$ 
in the $(z,t)$-plane for three different meron pair separations.
As the separation vanishes, the zero mode
goes over into the well-known instanton zero mode result.

Like instanton-antiinstanton pairs, however, 
a meron pair is not a strict minimum of the action,
since it has a weak attractive interaction, Eq.~(\ref{action}),
and under repeated relaxation will coalesce to form an
instanton. 
Nevertheless, just as it is important to sum all the quasi-stationary
instanton-antiinstanton configurations to obtain essential
nonperturbative physics~\cite{coleman}, 
meron pairs may be expected to
play an analogous role.
A precise framework for including quasi-stationary meron pairs is to
express the partition function as an integral over a suitable
collective variable $q$ with a constraint ${\cal Q}[A]$ as follows
\begin{eqnarray}
Z &=& \int\!\! DA \; \exp \left\{-S[A]\right\} = \int\!\! dq \; Z_q \ ,
\\
Z_q &=& \int\!\! DA \; \exp \left\{ -S[A] - \lambda 
\left( {\cal Q}[A]-q
\right)^2 \right\} \ .
\label{zq}
\end{eqnarray}
The meron pair is then a true minimum of the effective action in
Eq.~(\ref{zq}), allowing for a semiclassical treatment of $Z_q$.
We choose ${\cal Q}[A]$ to be the quadrupole moment $3z^2-x^2-y^2-t^2$ of
the topological charge.  

\begin{figure}[t]
\begin{center}
\leavevmode
\vspace*{-.5cm}
\epsfxsize=3.4in
\epsffile{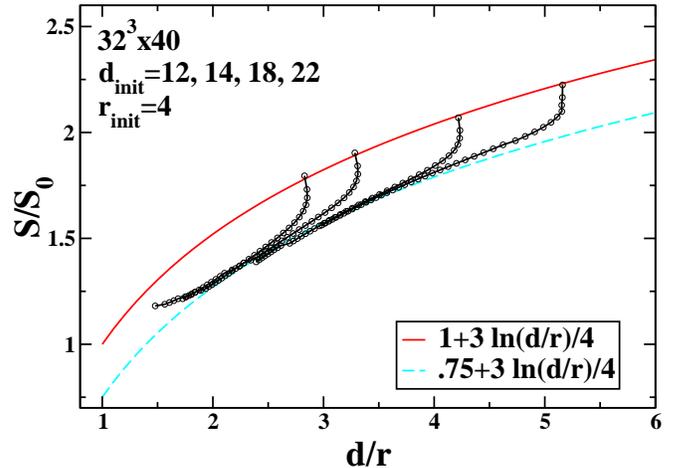}
\end{center}
\vspace{-.1cm}
\caption{\label{fig4} Action of a regulated meron pair 
as a function of $d/r$ for initial lattice configurations of $d_{\rm
init}=12$, 14, 18, 22 and $r_{\rm init}=4$ on a $32^3\times40$ lattice. 
The circles show cooling trajectories with each cooling step marked.}
\vspace{-.35cm}
\end{figure}

The criterion for a good collective variable $q$ of the system is that
the gradient in the direction of $q$ is small compared to the
curvature associated with all the quantum fluctuations.
In this case, we have an adiabatic limit in which 
relaxation of the unconstrained
meron pair slowly evolves through a sequence of
quasi-stationary solutions,
each of which is close to a corresponding stationary constrained solution. 
Detailed comparison of the quasi-stationary and constrained solutions
shows this adiabatic limit is well satisfied.
Therefore, we will present here the action of a
meron pair as it freely coalesces into an instanton.
To compare with the patched {\it Ansatz}, we need the meron separation $d$
and radii $r$ and $R$ of the lattice configuration.
We find these by first measuring the separation of the
two maxima in the action density $\Delta\equiv|M_{\rm I}-M_{\rm III}|$ 
(using cubic splines)
and their values (which are the same in the symmetric case,
denoted by $S_{\rm max}=48 /g^2 w^4$),
and then using Eq.~(\ref{maxima}) to give
\begin{equation}
d^2 = \Delta^2 + 4 w^2 \ ,
\qquad
\{r,R\} = \frac{2w d}{d\pm\Delta} \ .
\end{equation}
In Fig.~\ref{fig4}, we plot the total action of a meron pair as a function
of $d/r$ (solid line), which for the analytic case is given by
Eq.~(\ref{action}) with $R=d^2/r$. 
Also shown are cooling trajectories for 
four lattice configurations with different
initial meron pair separations.
Each case starts with a patched {\it Ansatz} of a given separation and the
lattice action agrees with that of the analytic {\it Ansatz}.

The first few cooling steps primarily decrease the action without
changing the collective variable (which is now effectively $d/r$), 
as observed above in Fig.~\ref{fig2}.
Further cooling gradually decreases the collective variable, tracing
out a new
logarithmic curve for the total action
(dashed line in Fig.~\ref{fig4}), 
which is about $0.25$ lower than the analytic case.
This curve represents the total action of the smooth adiabatic or
constrained lattice meron pair solutions.
The essential property of merons that could allow 
them to dominate the path integral and confine color charge
is the logarithmic interaction Eq.~(\ref{action})
which is weak enough to be dominated by the meron 
entropy~\cite{Callan:1977qs}.
Hence, the key physical result from Fig.~\ref{fig4} is the fact that
smooth adiabatic or constrained lattice meron pair solutions clearly
exhibit this logarithmic behavior.

In summary, we have found stationary meron pair solutions, shown that
they have a logarithmic interaction, and shown that they have
characteristic fermion zero modes localized about the individual
merons. 
Hence the presence and role of merons in numerical evaluations of the
QCD path integral should be investigated, and the study of fermionic
zero modes and cooling of gauge configurations provide possible tools
to do so.

\medskip

We would like to thank J.~Kuti and F.~Lenz for useful discussions.
This work was supported in part by the U.S. Department of Energy under
cooperative research agreement \#DF-FC02-94ER40818.


\vspace*{-.4cm}

\begin{references}
\vspace*{-1.5cm}

\bibitem{Chu:1994vi}
M.~C.~Chu, J.~M.~Grandy, S.~Huang and J.~W.~Negele,
Phys.\ Rev.\  {\bf D49}, 6039 (1994).
%
\bibitem{Schafer:1998wv}
T.~Sch\"afer and E.~V.~Shuryak,
Rev.\ Mod.\ Phys.\  {\bf 70}, 323 (1998).
%
\bibitem{Callan:1977qs}
C.~G.~Callan, R.~Dashen and D.~J.~Gross,
Phys.\ Lett.\  {\bf B66}, 375 (1977);
%
Phys.\ Rev.\  {\bf D17}, 2717 (1978);
%
Phys.\ Rev.\  {\bf D19}, 1826 (1979).
%
\bibitem{Chen:1999ct}
D.~Chen, R.~C.~Brower, J.~W.~Negele and E.~Shuryak,
Nucl.\ Phys.\ Proc.\ Suppl.\  {\bf 73}, 512 (1999).
%
\bibitem{Smit:1991vg}
J.~Smit and A.~van der Sijs,
Nucl.\ Phys.\  {\bf B355}, 603 (1991);
%
M.~N.~Chernodub and M.~I.~Polikarpov,
hep-th/9710205, and references therein.
%
\bibitem{Gonzalez-Arroyo:1996zy}
A.~Gonzalez-Arroyo and P.~Martinez,
Nucl.\ Phys.\  {\bf B459}, 337 (1996);
%
M.~Faber, J.~Greensite, S.~Olejnik and D.~Yamada,
hep-lat/9912002.
%
\bibitem{Polyakov:1977fu}
A.~M.~Polyakov,
Nucl.\ Phys.\  {\bf B120}, 429 (1977).
%
\bibitem{Actor:1979in}
A.~Actor,
Rev.\ Mod.\ Phys.\  {\bf 51}, 461 (1979).
%
\bibitem{Laughton:1980vv}
D.~G.~Laughton,
Can.\ J.\ Phys.\  {\bf 58}, 845, 859 (1980).
%
\bibitem{deAlfaro:1976qz}
V.~de Alfaro, S.~Fubini and G.~Furlan,
Phys.\ Lett.\  {\bf B65}, 163 (1976).
%
\bibitem{Jackiw:1977pu}
R.~Jackiw and C.~Rebbi,
Phys.\ Rev.\  {\bf D16}, 1052 (1977).
%
\bibitem{Konishi:1999re}
K.~Konishi and K.~Takenaga,
hep-th/9911097.
%
\bibitem{Ivanenko:1998nb}
T.~L.~Ivanenko and J.~W.~Negele,
Nucl.\ Phys.\ Proc.\ Suppl.\  {\bf 63}, 504 (1998).
%
T.~Ivanenko, MIT Ph.D., 1997.
%
\bibitem{GarciaPerez:1994ki}
M.~Garcia Perez, A.~Gonzalez-Arroyo, J.~Snippe and P.~van Baal,
Nucl.\ Phys.\  {\bf B413}, 535 (1994).
%
\bibitem{Berg:1981nw}
B.~Berg,
Phys.\ Lett.\  {\bf B104}, 475 (1981).
%
\bibitem{coleman}
S.~Coleman, {\it Aspects of Symmetry}, (Cambridge University Press, 1988).
\end{references}
\end{document}